\RequirePackage[2020-02-02]{latexrelease}
\documentclass[superscriptaddress,preprint]{revtex4}  
\usepackage{amsfonts}
\usepackage{amsmath}
\usepackage{cancel}
\usepackage{graphicx}
\usepackage{mathcomp}
\usepackage{color}
\usepackage{ulem}
\usepackage{xcolor} 
\usepackage{setspace}

\usepackage{array}
\usepackage{booktabs}
\usepackage[utf8]{inputenc}
\usepackage{overpic}
\usepackage{subfigure}
\usepackage{chemarrow}

\newif\iffigures
\figurestrue
\renewcommand{\refname}{REFERENCES}

\def\undertilde#1{\mathord{\vtop{\ialign{##\crcr
				$\hfil\displaystyle{#1}\hfil$\crcr\noalign{\kern1.5pt\nointerlineskip}
				$\hfil\widetilde{}\hfil$\crcr\noalign{\kern1.5pt}}}}}

\begin{document}
	\title{Kinetic investigation of the planar Multipole Resonance Probe under arbitrary pressure}
	
	\author{Chunjie Wang}
	\affiliation{Institute of Theoretical Electrical Engineering, Ruhr University Bochum, D-44780 Bochum, Germany}
	\author{Michael Friedrichs}
	\affiliation{Department of Electrical Power Engineering, South Westphalia University of Applied Science, 59494 Soest, Germany}
	\author{Jens Oberrath}
	\affiliation{Department of Electrical Power Engineering, South Westphalia University of Applied Science, 59494 Soest, Germany}
	\author{Ralf Peter Brinkmann}
	\affiliation{Institute of Theoretical Electrical Engineering, Ruhr University Bochum, D-44780 Bochum, Germany}

\date{\today}

\begin{abstract}

Active plasma resonance spectroscopy (APRS) refers to a class of plasma diagnostic methods that use the ability of plasma to resonate at or near the electron plasma frequency for diagnostic purposes. The planar multipole resonance probe (pMRP) is an optimized realization of APRS. It has a non-invasive structure and allows simultaneous measurement of the electron density, electron temperature, and electron-neutral collision frequency. Previous work has investigated the pMRP through the Drude model and collision-less kinetic model. The Drude model misses important kinetic effects such as collision-less kinetic damping. The collision-less kinetic model is able to capture pure kinetic effects. However, it is only applicable to low-pressure plasma. To further study the behavior of the pMRP, we develop a collisional kinetic model in this paper, which applies to arbitrary pressure. In this model, the kinetic equation is coupled to the Poisson equation under the electrostatic approximation. The real part of the general admittance is calculated to describe the spectral response of the probe-plasma system. Both collision-less kinetic damping and collisional damping appear in the spectrum. This model provides a possibility to calculate the electron density, electron temperature, and electron-neutral collision frequency from the measurements.

\vspace{3mm}

\end{abstract}

\maketitle
\newpage

\section{Introduction}

In the past decades, the industrial application of plasma technology has made remarkable progress \cite{Lieberman2005,Shul2011}. Plasma is used in many fields, such as etching, cleaning, and deposition. In the industrial process, plasma parameters, such as electron density and electron temperature, directly affect the quality of the product. It is very important to accurately control plasma parameters throughout the process.\par

Precise control of plasma parameters depends on reliable diagnostic techniques. Representative measurement methods include the Langmuir probe \cite{Merlino2007,Godyak2015}, optical emission spectroscopy \cite{Donnelly2004,Zhu2010}, and APRS probes \cite{Lapke2013,Oberrath2014,Kim2016,Pohle2018,Pohle2020,Schulz2014-1,Schulz2014-2,Friedrichs2018,Friedrichs2019,Wang2021,Pandey2014,Arshadi2017-1,Arshadi2017-2,Kim2019,Yeom2020,Yeom2021,Lapke2008,Oberrath2014-2,Oberrath2020,Fiebrandt2017,Oberrath2016,Oberrath2018}. APRS (active plasma resonance spectroscopy) refers to a class of plasma diagnostic methods that use the ability of plasma to resonate at or near the electron plasma frequency $\omega_\mathrm{pe}$ for diagnostic purposes \cite{Lapke2013,Oberrath2014}. At present, APRS probes are considered to be promising diagnostic methods. An attractive feature of APRS probes is that they are insensitive to additional dielectric deposition on the probe tip. The detailed advantages and disadvantages of APRS probes can be found in \cite{Kim2016,Pohle2020}.\par

Considering that the insertion of the probe can lead to plasma density deletion around the probe tip and its holder, non-invasive probes are preferred in the sensitive plasma process. In the past decade, several planar-type APRS probes have been developed, such as the planar multipole resonance probe (pMRP) \cite{Pohle2018,Pohle2020,Schulz2014-1,Schulz2014-2,Friedrichs2018,Friedrichs2019,Wang2021}, curling probe \cite{Pandey2014,Arshadi2017-1,Arshadi2017-2}, and flat cutoff probe \cite{Kim2019,Yeom2020,Yeom2021}. These probes can be flatly embedded into the chamber wall or chuck for minimally invasive process monitoring.\par

The pMRP is developed from the spherical multipole resonance probe (MRP) \cite{Lapke2008,Oberrath2014-2,Oberrath2020,Fiebrandt2017}. It consists of two semi-disc electrodes covered by a thin dielectric layer, which maintains a high degree of geometric and electrical symmetry. The pMRP allows simultaneous measurement of the electron density, electron temperature, and electron-neutral collision frequency. It is a very promising non-invasive APRS probe.\par

To evaluate plasma parameters from the spectra detected by the probe, mathematical models are very important. In \cite{Lapke2013} (Drude model) and \cite{Oberrath2014} (kinetic model), the generic properties of electrostatic APRS probes were investigated by the functional analytic approach. For any possible probe design, the spectral response function can be expressed as a matrix element of the resolvent of the dynamical operator. The Drude model is mathematically simple but physically limited. It only covers collisional damping. However, in the kinetic spectral response, residual damping appears in the vanishing pressure limit. This collision-less damping can only be interpreted as kinetic effects, which have been verified in further studies \cite{Oberrath2016,Oberrath2018,Oberrath2020,Wang2021}. A brief comparison of the Drude model and kinetic model is shown in Tab.~\ref{Drude and Kinetic}.\par

\begin{table}[!h]
	\caption{Comparison of the Drude model and kinetic model}
	\label{Drude and Kinetic}
	\centering 
	\renewcommand\arraystretch{1.4}
	\setlength{\tabcolsep}{2mm}
	\begin{tabular}{p{0.48\textwidth}p{0.48\textwidth}}
		\toprule[1pt]
		Drude model & Kinetic model\\ 
		\midrule[0.5pt]
		formulated in 3D space 
		& formulated in 3D3V space\\
		accessible to standard simulation tools 
		& inaccessible to standard simulation tools \\
		covers collisional damping
		& covers collision-less and collisional damping\\
		yields electron density and collision frequency
		& yields electron density, temperature, and collision frequency\\
		\bottomrule[1pt]
	\end{tabular}
\end{table}

Previous work has studied the pMRP with the Drude model \cite{Pohle2020,Pohle2018,Schulz2014-1,Schulz2014-2,Friedrichs2018,Friedrichs2019,Wang2021} and collision-less kinetic model \cite{Wang2021}. The collision-less kinetic model ignores all collisions, and it only applies to low-pressure plasma (a few $\mathrm{Pa}$). By comparing the spectrum of the Drude model and collision-less kinetic model, we found that both models can provide an accurate resonance frequency for calculating the electron density from the measured spectrum. But the collision-less kinetic model also yields the electron temperature, which is not available in the Drude model. As predicted in \cite{Oberrath2014}, the kinetic spectrum is obviously broadened by collision-less kinetic damping. This damping is non-negligible and even plays a dominant role in low-pressure plasma. However, the Drude model only includes collisional damping. This indicates that the collision frequency calculated by the Drude model is very inaccurate, especially in low-pressure plasma. The collision-less kinetic model only covers collision-less kinetic damping. As the gas pressure increases, the collisional damping gets stronger, which cannot always be ignored.\par

In this paper, we will present a collisional kinetic model that applies to arbitrary pressure. This collisional kinetic model is expected to cover both collision-less kinetic damping and collisional damping. As described in \cite{Oberrath2014}, we assume that 
$\omega_\mathrm{pe} \gtrsim \omega \gtrsim \nu \gg \nu_\mathrm{i} \approx \omega_\mathrm{pi} \gg \omega_\mathrm{g}$
(voltage frequency of the pMRP $\omega$, elastic collision frequency $\nu$, inelastic collision frequency $\nu_\mathrm{i}$, ion plasma frequency $\omega_\mathrm{pi}$, slow frequencies of all neutral gas phenomena $\omega_\mathrm{g}$). This assumption allows us to focus only on the dominant collision in electron dynamics: electron-neutral elastic collision. Obviously, this model is mathematically more complicated than the collision-less kinetic model due to the introduction of collisions. Next, we will take argon plasma as an example to introduce the collisional kinetic model and its corresponding kinetic spectra.\par

\pagebreak
\section{Static equilibrium}

As shown in Fig.~\ref{Ideal pMRP}, the idealized pMRP consists of two semi-disc electrodes $E_\pm$ with a radius of $R$. The electrodes are insulated from each other and from the grounded chamber wall. A thin dielectric layer with a thickness of $d$ covers the electrodes and chamber wall. As described in \cite{Friedrichs2018,Wang2021}, we assume that the chamber wall is infinite and the insulator is ignorable. A naturally oriented Cartesian coordinate system $(x,y,z)$ is used, which locates the dielectric layer in $-d<z<0$ and the plasma in $z>0$.\par

When the electrodes of the pMRP are grounded $\bar{\Phi}|_{E_{\pm}}=0$, a static equilibrium appears in front of the dielectric layer. In low-pressure plasma, this equilibrium behaves as a collision-less planar sheath \cite{Wang2021}, which follows the Bohm model \cite{Bohm1949}. To describe the static equilibrium under arbitrary pressure, a chemistry-free cold ion model is adopted: the equation of continuity expresses a constant ion flux $-\Psi_\mathrm{i}$ towards the dielectric
\begin{align}
n_\mathrm{i}v_\mathrm{i}=-\Psi_\mathrm{i},
\end{align}
and the equation of motion expresses the acceleration under the electric field $\bar{E}$ and the collisional friction in the neutral gas background
\begin{align}
m_\mathrm{i}v_\mathrm{i} \frac{\partial v_\mathrm{i}}{\partial z} =
e \bar{E}- \frac{\sqrt{v_\mathrm{T}^2+v_\mathrm{i}^2}}{\lambda_\mathrm{i}} m_\mathrm{i} v_\mathrm{i}
\end{align}
with the transition speed $v_\mathrm{T}=630 \,\mathrm{m/s}$ in the argon plasma. In terms of the friction force, an effective model is taken that combines the regimes of constant collision frequency $v_\mathrm{T}/\lambda_\mathrm{i}$ at low ion velocity and constant mean free path $\lambda_\mathrm{i}$ at high ion velocity \cite{Hornbeck1951}.
Assuming a constant electron temperature $T_\mathrm{e}$, the electron follows the Boltzmann relation
\begin{align}
T_\mathrm{e} \frac{\partial n_\mathrm{e}}{\partial z} = -e n_\mathrm{e} \bar{E}.
\end{align}
The Poisson equation relates the electric field to the charge density
\begin{align}
\epsilon_\mathrm{0} \frac{\partial \bar{E}}{\partial z} = e ( n_\mathrm{i}-n_\mathrm{e} )
\end{align}
with
\begin{align}
\bar{E}=-\frac{\partial \bar{\Phi}}{\partial z}.
\end{align}
\par

To write equations in dimensionless form, we use the following reference quantities: density $\hat{n}$, electron temperature $T_\mathrm{e}$, electron plasma frequency $\hat{\omega}_\mathrm{pe}=\sqrt{e^2 \hat{n}/(m_\mathrm{e} \epsilon_0)}$, Debye length $\hat{\lambda}_\mathrm{D}=\sqrt{\epsilon_0 T_\mathrm{e}/(e^2 \hat{n})}$. The normalization follows: 
$z \rightarrow \hat{\lambda}_\mathrm{D} z$, 
$t \rightarrow t/\hat{\omega}_\mathrm{pe}$, 
$\bar{\Phi} \rightarrow (T_\mathrm{e}/e) \bar{\Phi}$, 
$\bar{E} \rightarrow \sqrt {\hat{n} T_\mathrm{e}/\epsilon_0} \bar{E}$, 
$n_\mathrm{i} \rightarrow \hat{n} n_\mathrm{i}$, 
$n_\mathrm{e} \rightarrow \hat{n} n_\mathrm{e}$, 
$v_\mathrm{i} \rightarrow \sqrt{T_\mathrm{e}/m_\mathrm{i}} v_\mathrm{i}$, 
$v_\mathrm{e} \rightarrow \sqrt{T_\mathrm{e}/m_\mathrm{e}} v_\mathrm{e}$, 
$\Psi_\mathrm{i} \rightarrow \hat{n}\sqrt{T_\mathrm{e}/m_\mathrm{i}} \Psi_\mathrm{i}$, 
$v_\mathrm{T} \rightarrow \sqrt{T_\mathrm{e}/m_\mathrm{i}} v_\mathrm{T}$, 
$\lambda_\mathrm{i} \rightarrow \hat{\lambda}_\mathrm{D} \lambda_\mathrm{i}$.
In dimensionless form, the sheath model reads 
\begin{align}
n_\mathrm{i}v_\mathrm{i} &=-\Psi_\mathrm{i},\\
v_\mathrm{i} \frac{\partial v_\mathrm{i}}{\partial z}
&=\bar{E}-\frac{\sqrt{v_\mathrm{T}^2+v_\mathrm{i}^2}}{\lambda_\mathrm{i}} v_\mathrm{i},\\
\frac{\partial n_\mathrm{e}}{\partial z} &= -n_\mathrm{e} \bar{E},\\
\frac{\partial \bar{E}}{\partial z} & = n_\mathrm{i}-n_\mathrm{e},\\
\bar{E} & =-\frac{\partial \bar{\Phi}}{\partial z}.
\end{align}
Fig.~\ref{Sheath} depicts a floating sheath in an argon plasma at $10\,\mathrm{Pa}$. \par

\pagebreak
\section{Unperturbed trajectory}

The motion of the electron under the static potential $\bar{\Phi}$ is defined as the unperturbed trajectory. In \cite{Wang2021}, the collision-less kinetic model of the idealized pMRP was solved based on the unperturbed trajectory. This approach will also be adopted in this paper. As in \cite{Wang2021}, we introduce the coordinate system $(\varepsilon_z,\tau)$ based on the unperturbed trajectory. $\varepsilon_z$ is the electron's total energy in the $z$ direction
\begin{equation}
\varepsilon_z=\mathcal{E}_z \left(z,v_z\right) = \frac{1}{2} v_z^2 - \bar\Phi\left(z\right).
\end{equation}
$\tau$ describes the temporal parametrization of the unperturbed trajectory by selecting a turnaround ($\varepsilon_z<-\bar\Phi(0)$) or reflection ($\varepsilon_z>-\bar\Phi(0)$) as its reference point
\begin{equation}
\tau=\mathcal{T}\left(z,v_z\right) = \mathrm{sign}\left(v_z\right) \int_{z_\mathrm{min}}^z \frac{1}{\sqrt{v_z^2 - 2\bar\Phi\left(z\right) + 2\bar\Phi\left(z^\prime\right)}} \,\mathrm{d} z^\prime,
\end{equation}
with
\begin{equation}
z_\mathrm{min}=
\begin{cases}
\displaystyle{{\bar\Phi}^{-1}\left( -\frac{1}{2} v_z^2 + \bar\Phi\left(z\right) \right)} \quad  & \displaystyle{\frac{1}{2}v_z^2-\bar\Phi\left(z\right) < -\bar\Phi\left(0\right)} \\[2mm]
0    &\displaystyle{\frac{1}{2}v_z^2-\bar\Phi\left(z\right) \geq -\bar\Phi\left(0\right)}
\end{cases}.
\end{equation}
A coordinate transformation is then defined as
\begin{equation}
(z, v_z) \, \autorightleftharpoons{$\varepsilon_z=\mathcal{E}_z(z, v_z), \, \tau=\mathcal{T}(z, v_z)$}{$z=Z(\varepsilon_z,\tau), \, v_z=V_z(\varepsilon_z,\tau)$} \, (\varepsilon_z,\tau).
\end{equation}
More detailed descriptions of the unperturbed trajectory, coordinate system $(\varepsilon_z,\tau)$, and coordinate transformation $(z, v_z)\rightleftharpoons(\varepsilon_z,\tau)$ can be found in \cite{Wang2021}. Here, we define a new parameter $\varepsilon$ as the electron's total energy
\begin{align}
\varepsilon
=\mathcal{E}(z,\left|\vec{v}\right|)
=\frac{1}{2}\left|\vec{v}\right|^2-\bar{\Phi}(z).
\end{align}
Both $\varepsilon_z$ and $\varepsilon$ remain constant on the unperturbed trajectory. These parameters and coordinate transformation will be used in later calculations.\par

\pagebreak

\section{Kinetic model of the probe-plasma system}

A static planar sheath appears in front of the pMRP when its electrodes are grounded. During the measurement process, RF voltages are applied to the electrodes
\begin{align}
\Phi|_{E_{\pm}}=\pm \hat{V} \cos (\omega t),
\end{align}	
thereby generating a dynamic perturbation in the plasma around the probe. 
Since $\omega \gg \omega_\mathrm{pi}$, the ions dynamics under the RF electric field are neglected. In electron dynamics, we introduce the electron-neutral elastic collision. In the limit of $m_\mathrm{e}/m_\mathrm{N} \rightarrow 0$, the neutral particle is considered as an immobile scattering center, and it finally yields an angle and velocity independent differential collision frequency by assuming the hard-sphere collision. The electron distribution function $f(\vec{r},\vec{v},t)$ thus follows 
\begin{align}
\frac{\partial f}{\partial t}+\vec{v}\cdot\nabla_r f+\nabla\Phi\cdot\nabla_v f 
=\frac{\nu}{4 \pi}\int_{\Omega} f(\vec{r},\left|\vec{v}\right|\vec{e},t)\mathrm{d}\Omega -\nu f.
\label{Boltzmann}
\end{align}
As in \cite{Friedrichs2018,Wang2021}, the potential follows the Poisson equation under the electrostatic approximation
\begin{align}
-\nabla \cdot ( \epsilon_\mathrm{r} \nabla \Phi )=
\begin{cases}
0                                  &\text{Dielectric} \\
\displaystyle{n_\mathrm{i}-\int f \, \mathrm{d}^3 v}   \quad &\text{Plasma} 
\end{cases}
\label{Poisson}
\end{align}
with $\epsilon_\mathrm{r}=\epsilon_\mathrm{D}$ in the dielectric layer and $\epsilon_\mathrm{r}=1$ in the plasma.\par

Under a small perturbation, the linear response theory applies \cite{Krall1973,Buckley1966}: $f$ and $\Phi$ can be split into an equilibrium value and a small perturbation
\begin{align}
f(\vec{r},\vec{v},t)&=\bar f(z,\vec{v}) (1+\delta\!f(\vec{r},\vec{v},t)),
\label{linearization-f}\\
\Phi(\vec{r},t)&=\bar \Phi(z)+\delta \Phi(\vec{r},t),
\label{linearization-Phi}
\end{align}
in which
\begin{align}
\bar f (z,\vec{v})=f_\mathrm{M} (z,\vec{v})
&=\frac{1}{(2 \pi)^{3/2}} 
\exp\left(-\frac{1}{2}(v_x^2+v_y^2+v_z^2)+\bar{\Phi}(z)\right),\\
\left| \delta\!f \right| & \ll 1,\\
\left| \delta\Phi \right| & \ll 1.
\end{align}
Substituting \eqref{linearization-f} and \eqref{linearization-Phi} into \eqref{Boltzmann} and \eqref{Poisson}, the linearized equations read
\begin{align}
\frac{\partial \delta\!f}{\partial t}+\vec{v} \cdot \nabla_r \delta\!f -\vec{v} \cdot \nabla \delta\Phi + \bar\Phi^\prime (z) \frac{\partial \delta\!f}{\partial v_z}
&=\frac{\nu}{4 \pi}\int_{\Omega} \delta\!f(\vec{r},\left|\vec{v}\right|\vec{e},t)\mathrm{d}\Omega-\nu \delta\!f, 
\label{f1-perturbation}\\
-\nabla \cdot (\epsilon_\mathrm{r} \nabla \delta\Phi )
&=
\begin{cases}
0                                                 &\text{Dielectric} \\
\displaystyle{-\int \bar f\,\delta\!f\,{\mathrm{d}^3 v}}  \quad &\text{Plasma}  
\end{cases}.
\label{phi1-perturbation}
\end{align}
As in \cite{Wang2021}, we assume that the perturbation terms are time-harmonic
\begin{align}
\delta\!f (\vec{r},\vec{v},t)
&=\mathrm{Re}\left[\delta\!\tilde{f} (\vec{r},\vec{v}) \exp(\mathrm{i} \omega t )\right],\\
\delta\Phi (\vec{r},t)
&=\mathrm{Re}\left[\delta\tilde{\Phi} (\vec{r}) \exp(\mathrm{i} \omega t)\right].
\end{align}
When the electrodes of the pMRP are grounded, the equilibrium distribution varies only in the $z$ direction. Therefore, we apply the Fourier transform
\begin{align}
\delta\!\underline{\tilde{f}}(k_x,k_y,z,\vec{v})
&=\int_{-\infty}^{\infty} \int_{-\infty}^{\infty} \delta\!\tilde{f}(\vec{r},\vec{v}) \exp(\mathrm{i} (k_x x+k_y y)) \mathrm{d}x \mathrm{d}y,\\
\delta \underline{\tilde{\Phi}}(k_x,k_y,z)
&=\int_{-\infty}^{\infty} \int_{-\infty}^{\infty} \delta\tilde{\Phi}(\vec{r}) \exp(\mathrm{i}(k_x x+k_y y)) \mathrm{d}x \mathrm{d}y.
\end{align}
Now, the uniformity of the equilibrium on $x$ and $y$ is transferred to $k_x$ and $k_y$ by the Fourier transform. In the following calculations, $k_x$ and $k_y$ are temporarily neglected in $\delta\!\underline{\tilde{f}}$ and $\delta \underline{\tilde{\Phi}}$ before performing the inverse Fourier transform. Through the time-harmonic assumption and the Fourier transform, the partial derivatives in \eqref{f1-perturbation} and \eqref{phi1-perturbation} are simplified
\begin{align}
\frac{\partial\ }{\partial t} &\rightarrow \mathrm{i}  \omega,\\
\frac{\partial\ \; }{\partial x} &\rightarrow -\mathrm{i}  k_x,\\
\frac{\partial\ \; }{\partial y} &\rightarrow -\mathrm{i}  k_y.
\end{align}
The linearized equations \eqref{f1-perturbation} and \eqref{phi1-perturbation} become
\begin{gather}
\begin{aligned} 
(\nu+\mathrm{i}(\omega-k_x v_x-k_y v_y)) \delta\!\underline{\tilde{f}} 
+ v_z \frac{\partial \delta\!\underline{\tilde{f}}}{\partial z} 
+&
\bar\Phi^\prime(z) \frac{\partial \delta\!\underline{\tilde{f}}}{\partial v_z} 
+ \mathrm{i}(k_x v_x+k_y v_y)\delta\underline{\tilde{\Phi}} 
- v_z \frac{\partial \delta\underline{\tilde{\Phi}}}{\partial z}
\\
=&\frac{\nu}{4 \pi}\int_{\Omega}\delta\!\underline{\tilde{f}} (z,\left|\vec{v}\right|\vec{e})\mathrm{d}\Omega,
\label{kinetic-fourier1}
\end{aligned}
\\
\qquad
(k_x^2+k_y^2)\delta\underline{\tilde{\Phi}}-\frac{\partial^2 \delta\underline{\tilde{\Phi}}}{\partial z^2}=
\begin{cases}
0             &\text{Dielectric} \\
\displaystyle{-\int \bar f \,\delta\!\underline{\tilde{f}} \,\mathrm{d}^3 v}  \quad &\text{Plasma} 
\end{cases}.
\label{Poisson-fourier1}
\end{gather}
\par

The kinetic equation \eqref{kinetic-fourier1} is very complicated because of the presence of the integral in the collision term. Here, we define a new function
\begin{align}
\delta\!\underline{\tilde{g}}(z,\varepsilon)=\frac{1}{4 \pi}\int_{\Omega}\delta\!\underline{\tilde{f}}\left(z,\sqrt{2\left(\varepsilon+\bar{\Phi}(z)\right)}\,\vec{e}\right)\mathrm{d}\Omega,
\label{deltag1}
\end{align}
so
\begin{align}
\frac{1}{4 \pi}\int_{\Omega}\delta\!\underline{\tilde{f}} (z,\left|\vec{v}\right|\vec{e})\mathrm{d}\Omega=\delta\!\underline{\tilde{g}}\left(z,\frac{1}{2}\left|\vec{v}\right|^2-\bar{\Phi}(z)\right).
\label{deltag2}
\end{align}
Substituting \eqref{deltag2} into \eqref{kinetic-fourier1}, and then transforming $(z,v_z)$ into $(\varepsilon_z,\tau)$, the kinetic equation is simplified into
\begin{align}
\begin{gathered}
(\nu + \mathrm{i} (\omega - k_x v_x - k_y v_y) ) \delta\!\underline{\tilde{f}} ( Z(\varepsilon_z,\tau ),v_x,v_y,V_z(\varepsilon_z,\tau ) )+\frac{\partial \delta\!\underline{\tilde{f}}( Z(\varepsilon_z,\tau ),v_x,v_y,V_z(\varepsilon_z,\tau ) )}{\partial \tau}\\
+ \mathrm{i} (k_x v_x+k_y v_y ) \delta\underline{\tilde{\Phi}}( Z(\varepsilon_z,\tau))
-\frac{\partial \delta\underline{\tilde{\Phi}}( Z(\varepsilon_z,\tau ))}{\partial \tau} 
=\nu \delta\!\underline{\tilde{g}}\left(Z(\varepsilon_z,\tau ),\frac{1}{2}( v_x^2+v_y^2 )+\varepsilon_z \right).
\end{gathered}
\end{align}
By integrating along the unperturbed trajectory \cite{Buckley1966}, we get the solution of the kinetic equation
\begin{align}
\begin{aligned}
& \delta\!\underline{\tilde{f}} ( Z(\varepsilon_z,\tau ),v_x,v_y,V_z (\varepsilon_z,\tau ) )\\
=& \delta\underline{\tilde{\Phi}}( Z(\varepsilon_z,\tau ) )-( \nu+\mathrm{i}\omega ) \int_{-\infty}^{\tau} \exp((\nu+\mathrm{i} (\omega - k_x v_x - k_y v_y))(\tau^{\prime}-\tau) ) \delta\underline{\tilde{\Phi}}(Z(\varepsilon_z,\tau^{\prime}))\, {\mathrm{d} \tau^{\prime}}\\
& + \nu \int_{-\infty}^{\tau} \exp((\nu+\mathrm{i} (\omega - k_x v_x - k_y v_y))(\tau^{\prime}-\tau) ) \,\delta\underline{\tilde{g}} ( Z(\varepsilon_z,\tau^{\prime}),\frac{1}{2}( v_x^2+v_y^2 )+\varepsilon_z ) \mathrm{d} \tau^{\prime}
\end{aligned},
\label{solution of kinetic equation}
\end{align} 
where all the perturbation is accumulated along the unperturbed trajectory through the integral over $\tau^{\prime}$ from  $-\infty$ to $\tau$.\par

Substituting \eqref{solution of kinetic equation} into \eqref{deltag1} (Section \ref{section delta g}), the definition of $\delta\!\underline{\tilde{g}}\left(z,\varepsilon \right)$ yields
\begin{align}
\delta\underline{\tilde{g}}(z,\varepsilon)
-\delta\underline{\tilde{\Phi}}(z) 
= \int_0^{\infty} K_0 (\nu+\mathrm{i}\omega,k,z,z^{\prime},\varepsilon) 
\left(\nu \, \delta\underline{\tilde{g}}(z^{\prime},\varepsilon)
-(\nu+\mathrm{i} \omega) \delta\underline{\tilde{\Phi}}(z^{\prime})\right) 
\mathrm{d} z^{\prime},
\label{deltag3}
\end{align}
where $k=\sqrt{k_x^2+k_y^2}$ and $K_0$ is given in Section \ref{section delta g}.
Substituting \eqref{solution of kinetic equation} into \eqref{Poisson-fourier1} (Section \ref{section Poisson equation}), the Poisson equation becomes
\begin{align}
\begin{aligned}
&k^2 \delta\underline{\tilde{\Phi}}(z)-\frac{\partial^2 \delta\underline{\tilde{\Phi}}(z)}{\partial z^2}\\
=&
\begin{cases}
0      &\text{Dielectric} \\[2mm]
\begin{aligned}
&-\exp\left(\bar\Phi(z)\right)
\delta\underline{\tilde{\Phi}}(z)
+(\nu+\mathrm{i}\omega) \int_{0}^{+\infty} K_1(\nu+\mathrm{i}\omega,k,z,z^{\prime}) \delta\underline{\tilde{\Phi}} (z^{\prime}) \mathrm{d} z^{\prime}\\
&-\nu \int_{-\bar{\Phi}\left(z\right)}^{+\infty} \int_{0}^{+\infty} 
K_2(\nu+\mathrm{i}\omega,k,z,z^{\prime},\varepsilon)
\delta\underline{\tilde{g}} (z^{\prime},\varepsilon) 
\mathrm{d} z^{\prime} \mathrm{d} \varepsilon
\end{aligned}  &\text{Plasma} 
\end{cases},
\end{aligned}
\label{Poisson-fourier2}
\end{align}
where $K_1$ and $K_2$ are given in Section \ref{section Poisson equation}. Due to the uniformity of the static equilibrium in the $x$ and $y$ direction, $k=\sqrt{k_x^2+k_y^2}$ eventually appears in \eqref{deltag3} and \eqref{Poisson-fourier2}. In the Fourier space, the boundary conditions read
\begin{align}
\delta\underline{\tilde{\Phi}}(-d)& =\hat{V} \iint_{E_\pm}  \mathrm{sign} (y) \exp(\mathrm{i}(k_x x+k_y y))\mathrm{d}x \mathrm{d}y,\\
\delta\underline{\tilde{\Phi}}(0_-)& =\delta\underline{\tilde{\Phi}}(0_+),\\
\epsilon_\mathrm{D} \delta\underline{\tilde{\Phi}}^{\prime}(0_-) &=\delta\underline{\tilde{\Phi}}^{\prime}(0_+),\\
\delta\underline{\tilde{\Phi}}(+\infty)& =0.
\end{align}
In the plasma domain, the Poisson equation finally becomes a complicated integral-differential equation \eqref{Poisson-fourier2}, which is coupled to a complicated integral equation \eqref{deltag3}. Obviously, the analytical solution of the potential is inaccessible.\par

\pagebreak

\section{Numerical solution of the Poisson equation}

To solve the Poisson equation \eqref{Poisson-fourier2} numerically, we define
\begin{align}
z_n =
\begin{cases}
n \Delta z   
&
N_1 \leq n < 0 \\
\bar{\Phi}^{-1}(\bar{\Phi}(0)+n \Delta \varepsilon)     
&
0 \leq n \leq N_2
\end{cases}
\label{definition of zn}
\end{align}
and
\begin{align}
\varepsilon_n &=-\bar{\Phi}(0) + (n-N_2) \Delta \varepsilon  
\quad 
0 \leq n \leq N_3,
\end{align}
where $\Delta z$ is the grid size of $z$ when $z \leq 0$, $\Delta \varepsilon$ is the grid size of $\varepsilon$, and $N_1<0<N_2<N_3$. In the definition of $z_n$ \eqref{definition of zn}, an even spatial grid $\Delta z$ is adopted in the dielectric domain
\begin{align}
z_{N_1+1} & =-d,\\
z_{0} & =0,
\end{align}
and an even potential grid $\Delta \varepsilon$ is adopted in the plasma domain
\begin{align}
\bar{\Phi}(z_{n+1})-\bar{\Phi}(z_{n})=\Delta \varepsilon  \qquad 0 \leq n<N_2.
\end{align}
Here, we assume that the perturbation of the potential is negligible when $z>z_{N_2}$. 
Then we employ
\begin{align}
\delta\underline{\tilde{g}}(z,\varepsilon)
&=\sum_{n=1}^{N_3} \,
\sum_{m=1}^{N_2}
\delta\underline{\tilde{g}}_{m,n} \alpha_{m,n}(z,\varepsilon),
\\
\delta\underline{\tilde{\Phi}}(z) 
=&\sum_{m=N_1+1}^{N_2} 
\delta\underline{\tilde{\Phi}}_m \beta_m(z),
\end{align}
in which
\begin{align}
\alpha_{m,n}(z,\varepsilon)&=
\begin{cases}
1  & z_{m-1}<z \leq z_m \ \text{and}\ \varepsilon_{n-1}<\varepsilon \leq \varepsilon_n\\
0  & \text{Others}
\end{cases},
\\
\beta_m(z)&=
\begin{cases}
1  & z_{m-1}<z \leq z_m\\
0  & \text{Others}
\end{cases}.
\end{align}
\par

\begin{table}[!h]
	\caption{Illustration of the matrices}
	\centering 
	\renewcommand\arraystretch{1.4}
	\setlength\tabcolsep{2mm}
	\begin{tabular}{p{0.09\textwidth}p{0.56\textwidth}p{0.28\textwidth}}
		\specialrule{1pt}{0pt}{0pt}
		Matrix & Description & Dimension\\ 
		\specialrule{0.5pt}{0pt}{2pt} 
		$\mathbf{A}_n$
		& related to $K_0$ in \eqref{deltag3}
		& $\min{(n,N_2)} \times \min{(n,N_2)}$
		\\
		$\mathbf{B}_n$
		& related to $K_2$ in \eqref{Poisson-fourier2}
		& $\min{(n,N_2)} \times \min{(n,N_2)}$
		\\
		$\mathbf{C}$
		& related to \eqref{Poisson-fourier2}
		& $(N_2-N_1) \times (N_2-N_1)$
		\\
		\specialrule{0pt}{0pt}{2pt} 
		$\mathbf{d}$
		& $\left( 
		\delta\underline{\tilde{\Phi}}(-d) \quad 
		0 \ \cdots \ 0 \right)^{T}$
		& $(N_2-N_1) \times 1$
		\\
		\specialrule{0pt}{2pt}{0pt} 				
		$\mathbf{I}_n$
		& unit matrix
		& $\min{(n,N_2)} \times \min{(n,N_2)}$
		\\
		$\mathbf{0}_n$
		& zero matrix
		& $(N_2-N_1-\min{(n,N_2)}) \times 1$
		\\
		\specialrule{0pt}{0pt}{2pt}
		$\mathbf{\delta\underline{\tilde{g}}}_n$
		& $\left( 
		\delta\underline{\tilde{g}}_{\max{(N_2-n,0)}+1,n} \quad 
		\delta\underline{\tilde{g}}_{\max{(N_2-n,0)}+2,n} \ \cdots \ 
		\delta\underline{\tilde{g}}_{N_2,n} \right)^{T}$
		& $\min{(n,N_2)} \times 1$
		\\
		\specialrule{0pt}{2pt}{2pt}
		$\mathbf{\delta\underline{\tilde{\Phi}}}_n$
		&$\left( 
		\delta\underline{\tilde{\Phi}}_{\max{(N_2-n,0)}+1} \quad 
		\delta\underline{\tilde{\Phi}}_{\max{(N_2-n,0)}+2} \ \cdots \ 
		\delta\underline{\tilde{\Phi}}_{N_2} \right)^{T}$
		& $\min{(n,N_2)} \times 1$
		\\
		\specialrule{0pt}{2pt}{2pt}
		$\mathbf{\delta\underline{\tilde{\Phi}}}$
		&$\left(
		\delta\underline{\tilde{\Phi}}_{N_1+1} \quad 
		\delta\underline{\tilde{\Phi}}_{N_1+2} \ \cdots \ 
		\delta\underline{\tilde{\Phi}}_{N_2} \right)^{T}$
		& $(N_2-N_1) \times 1$
		\\
		\specialrule{1pt}{2pt}{0pt}
	\end{tabular}
\end{table}

For $\varepsilon=\varepsilon_n$ ($n=1,2,\cdots,N_3$), \eqref{deltag3} can be numerically discretized into a matrix equation
\begin{align}
\left(\mathbf{I}_n-\nu \mathbf{A}_n\right)
\mathbf{\delta\underline{\tilde{g}}}_n
=\left(\mathbf{I}_n-(\nu+\mathrm{i}\omega) \mathbf{A}_n \right)
\mathbf{\delta\underline{\tilde{\Phi}}}_n,
\label{ME g phi}
\end{align}
thus
\begin{align}
\mathbf{\delta\underline{\tilde{g}}}_n
=\left(\mathbf{I}_n-\nu \mathbf{A}_n\right)^{-1}
\left(\mathbf{I}_n-(\nu+\mathrm{i}\omega) \mathbf{A}_n \right)
\mathbf{\delta\underline{\tilde{\Phi}}}_n. 
\label{solution of g}
\end{align}
The Poisson equation \eqref{Poisson-fourier2} and the corresponding boundary conditions can be discretized into
\begin{equation}
\nu
\sum_{n=1}^{N_3} 
\left[\begin{array}{c}
\mathbf{0}_n\\
\mathbf{B}_n
\mathbf{\delta\underline{\tilde{g}}}_n
\end{array}\right] 
+\mathbf{C} \, \mathbf{\delta\underline{\tilde{\Phi}}}
=\mathbf{d}.
\label{ME-Poisson}
\end{equation}
Substituting \eqref{solution of g} into \eqref{ME-Poisson}, we obtain
\begin{equation}
\nu \sum_{n=1}^{N_3}
\left[\begin{array}{c}
\mathbf{0}_n\\
\mathbf{B}_n
\left(\mathbf{I}_n-\nu \mathbf{A}_n\right)^{-1}
\left(\mathbf{I}_n-(\nu+\mathrm{i}\omega) \mathbf{A}_n\right)
\mathbf{\delta\underline{\tilde{\Phi}}}_n
\end{array}\right]
+\mathbf{C} \, \mathbf{\delta\underline{\tilde{\Phi}}}
=\mathbf{d},
\end{equation}
which yields the numerical solution of the potential. Afterwards, we can calculate the real part of the general complex admittance $\mathrm{Re}[Y\left(\omega\right)]$. The detailed derivation of $\mathrm{Re}[Y\left(\omega\right)]$ can be found in \cite{Wang2021}.
\par

\pagebreak

\section{Spectral response of the probe-plasma system}

In \cite{Wang2021}, we have studied the behavior of the idealized pMRP in low-pressure plasma using a collision-less kinetic model. This model is able to capture pure collision-less kinetic damping, which obviously broadens the spectrum of the idealized pMRP. In the previous sections, we present a collisional kinetic model. This model takes into account the electron-neutral elastic collision and applies to arbitrary pressure. In this section, we will further investigate the kinetic spectrum of the idealized pMRP with a focus on collisional damping. The same idealized pMRP as in \cite{Wang2021} is chosen: electrode radius $R=5\,\mathrm{mm}$, dielectric thickness $d=0.04 \,\mathrm{mm}$, dielectric relative permittivity $\epsilon_\mathrm{D}=4.82$. The kinetic spectra are calculated when this probe monitors an argon plasma as shown in Fig.~\ref{Sheath}.\par

Fig.~\ref{Spectra} presents the kinetic spectra of the idealized pMRP, represented by the real part of its general complex admittance $\mathrm{Re}[Y\left(\omega\right)]$. When the electron-neutral collision frequency $\nu=0$, the spectral resonance and its broadening by pure kinetic effects are clearly visible. $\mathrm{Re}[Y\left(\omega\right)]$ reaches a maximum value of $2.43\,\mathrm{mS}$ at $0.46\,\hat{\omega}_\mathrm{pe}$ $(\hat{\omega}_\mathrm{pe}=5.64\,\mathrm{GHz})$, and the half-width of the resonance peak is $0.12\,\hat{\omega}_\mathrm{pe}$. As described in \cite{Oberrath2014}, the probe generates kinetic free energy. This energy can be transported by electrons from the probe to such a large distance that the probe cannot detect it. The loss of kinetic free energy will cause collision-less damping in the spectrum. When $\nu$ increases from $0$ to $0.2\hat{\omega}_\mathrm{pe}$, the spectrum exhibits almost the same resonance frequency. But the resonance peak becomes lower and the half-width becomes higher (Fig.~\ref{Spectra12}), which indicates the presence of stronger collisional damping. Hence, we can conclude that this collisional kinetic model is able to cover both collision-less damping and collisional damping.\par

The collisional kinetic model allows further study on the behavior of the idealized pMRP under arbitrary pressure. It offers the possibility to calculate the electron density, electron temperature, and electron-neutral collision frequency from the measured spectrum. However, it’s difficult to describe the relationship between the kinetic spectrum and plasma parameters with specific equations. One possible way is to build a spectral database from parameter studies at different $n_\mathrm{e}$, $T_\mathrm{e}$, and $\nu$. By comparing the measured spectrum with the spectral database, plasma parameters can be evaluated.\par

\pagebreak

\section{Conclusion and outlook}

To investigate the behavior of an idealized pMRP, we develop a collisional kinetic model which applies to arbitrary pressure. A static planar sheath appears in front of the pMRP when its electrodes are grounded. As the electrodes are applied with RF voltages, a dynamic perturbation will be generated around the probe. Under a small perturbation, the linearized kinetic model, including the kinetic equation and Poisson equation, is employed to study the kinetic spectral response of the probe-plasma system. Considering the planar geometry of the pMRP, we perform the Fourier transform in the directions parallel to the probe ($x$ and $y$). The formulas are then derived and solved in the Fourier space. This approach shows high superiority in planar geometry, and we can expect it to be applied to other planar-type APRS probes.\par

The spectral response of the idealized pMRP is expressed by the real part of the general complex admittance. When the electron-neutral collision frequency $\nu=0$, the spectral resonance and its broadening by pure kinetic effects are clearly visible. As $\nu$ increases from $0$ to $0.2\hat{\omega}_\mathrm{pe}$, stronger collisional damping appears in the kinetic spectrum. This collisional kinetic model covers both collision-less kinetic damping and collisional damping. It yields the electron density, electron temperature, and electron-neutral collision frequency.\par

In \cite{Friedrichs2019}, Friedrichs \textit{et al} compared the pMRP spectra of the analytic approach, electrostatic simulation, and electromagnetic simulation. They found that the main difference between the spectra of the idealized pMRP and the real pMRP is caused by the thickness of the insulator between the electrodes and the chamber wall. So far, our team has developed three analytic models: the Drude model \cite{Friedrichs2018}, the collision-less kinetic model \cite{Wang2021}, and the collisional kinetic model (this paper). But all these analytic models are based on the idealized pMRP geometry. Therefore, further work is needed to optimize these analytic models by taking into account the influence of insulator thickness, which will bring certain challenges to the potential calculation.\par

\pagebreak

\section{Acknowledgments}

The authors gratefully acknowledge the financial support by Deutsche Forschungsgemeinschaft (DFG) via the project DFG 360750908. Gratitude is expressed to the MRP-Team at Ruhr University Bochum.

\renewcommand\refname{\fontsize{11pt}{13pt} \selectfont References \\ \vspace*{-22pt}}
\begin{spacing}{1} 
	
\end{spacing}

\pagebreak

\section{Appendix A. Definition of $\delta\!\underline{\tilde{g}} \left(z,\varepsilon \right)$}
\label{section delta g}
\setcounter{equation}{0}
\renewcommand{\theequation}{A.\arabic{equation}}

Defining
\begin{align}
\delta\underline{\tilde{h}} \left( z,\varepsilon \right)=
\nu \delta\underline{\tilde{g}} \left( z,\varepsilon \right)
-\left( \nu+\mathrm{i}\omega \right) \delta\underline{\tilde{\Phi}}\left(z\right),
\end{align}
the solution of kinetic equation \eqref{solution of kinetic equation} becomes
\begin{align}
\begin{gathered}
\delta\!\underline{\tilde{f}} \left( Z\left(\varepsilon_z,\tau \right),v_x,v_y,V_z \left(\varepsilon_z,\tau \right) \right)
= \delta\underline{\tilde{\Phi}}\left( Z\left(\varepsilon_z,\tau \right) \right) \\
+ \int_{-\infty}^{\tau} \exp\left(\left(\nu+\mathrm{i} \left(\omega - k_x v_x - k_y v_y\right)\right)\left(\tau^{\prime}-\tau\right) \right) \,\delta\underline{\tilde{h}} \left( Z\left(\varepsilon_z,\tau^{\prime}\right),\frac{1}{2}\left( v_x^2+v_y^2 \right)+\varepsilon_z \right) \mathrm{d} \tau^{\prime}.
\end{gathered}
\label{skA}
\end{align}
Substituting \eqref{skA} into the definition of $\delta\!\underline{\tilde{g}} \left(z,\varepsilon \right)$ \eqref{deltag1}, the first part gives
\begin{align}
\frac{1}{4 \pi} \int_{\Omega} \int_0^{+\infty} \delta\underline{\tilde{\Phi}}\left( Z(\varepsilon_z,\tau) \right) \delta(Z(\varepsilon_z,\tau)-z) \mathrm{d}Z \, \mathrm{d}\Omega 
=\delta\underline{\tilde{\Phi}}(z),
\end{align}
and the second part gives
\begin{align}
\begin{aligned}
& \frac{1}{4 \pi} 
\int_{\Omega} \int_0^{+\infty} \int_0^{+\infty} \int_{-\infty}^{\tau}
\exp\left(\left(\nu+\mathrm{i} \left(\omega - k_x v_x - k_y v_y \right) \right) \left(\tau^{\prime}-\tau \right) \right) \\
& \delta\underline{\tilde{h}} \left( Z\left(\varepsilon_z,\tau^{\prime}\right),\frac{1}{2}\left( v_x^2+v_y^2 \right)+\varepsilon_z \right) 
\delta\left(Z \left(\varepsilon_z,\tau \right)-z \right) \\
& \delta \left(\sqrt{v_x^2+v_y^2+2\left(\varepsilon_z+\bar \Phi \left(z\right)\right)}-\sqrt{2\left(\varepsilon+\bar \Phi \left(z\right)\right)}\right) 
\mathrm{d} \tau^{\prime} \, \mathrm{d}Z \, \mathrm{d} V \, \mathrm{d}\Omega  \\
=&\frac{1}{4\sqrt{\varepsilon+\bar \Phi \left(z\right)}} 
\int_{0}^{\sqrt{2\left(\varepsilon+\bar \Phi \left(z\right)\right)}}  
\int_{-\bar\Phi\left(z\right)}^{\varepsilon} 
\frac{1}{\sqrt{\varepsilon_z+\bar \Phi \left(z\right)}}  
\int_{-\infty}^{+\infty}  \int_{-\infty}^{\tau} \exp\left(\left(\nu+\mathrm{i}\omega\right)\left(\tau^{\prime}-\tau\right)\right) \\
&J_0\left( k v_{xy} \left(\tau^{\prime}-\tau\right) \right)
\delta\underline{\tilde{h}} \left( Z\left(\varepsilon_z,\tau^{\prime}\right),\frac{1}{2}v_{xy}^2+\varepsilon_z \right) \\
&\left(\delta \left(\tau-\mathcal{T}\left(z, \sqrt{2\left( \varepsilon_z + \bar \Phi \left(z\right) \right)}\right)\right)+\delta\left(\tau-\mathcal{T}\left(z,-\sqrt{2\left( \varepsilon_z + \bar \Phi \left(z\right) \right)}\right)\right)\right)\\
& \delta\left(v_{xy}-\sqrt{2\left(\varepsilon-\varepsilon_z\right)}\right)
\mathrm{d} \tau^{\prime} \, \mathrm{d}\tau \, \mathrm{d} \varepsilon_z
\, \mathrm{d}v_{xy}\\
=&\int_0^{\infty} K_0 (\nu+\mathrm{i}\omega,k,z,z^{\prime},\varepsilon) 
\delta\underline{\tilde{h}}(z^{\prime},\varepsilon)
\mathrm{d} z^{\prime}
\end{aligned},
\end{align}

When $z^{\prime} \leq \bar{\Phi}^{-1}(-\varepsilon)$,
\begin{align}
K_0 (\nu+\mathrm{i}\omega,k,z,z^{\prime},\varepsilon)=0;
\end{align}
when $\bar{\Phi}^{-1}(-\varepsilon) < z^{\prime} \leq z$,
\begin{align}
\begin{aligned}
&K_0 (\nu+\mathrm{i}\omega,k,z,z^{\prime},\varepsilon)\\
=&
\int_{-\bar\Phi\left(z^{\prime}\right)}^{\varepsilon} 
\frac{1}{4\sqrt{
2\left(\varepsilon+\bar \Phi \left(z\right)\right)
\left(\varepsilon_z+\bar \Phi \left(z\right)\right)
\left(\varepsilon_z+\bar \Phi \left(z^{\prime}\right)\right)}}\\  
&\left(\exp{\left(\left(\nu+\mathrm i \omega\right)
\left(
\mathcal{T}\left(z^{\prime},\sqrt{2\left( \varepsilon_z + \bar \Phi \left(z^{\prime}\right) \right)}\right)-
\mathcal{T}\left(z,\sqrt{2\left( \varepsilon_z + \bar \Phi \left(z\right) \right)}\right) \right)\right)}\right.\\
&J_0\left(k \sqrt{2\left(\varepsilon-\varepsilon_z\right)}
\left(
\mathcal{T}\left(z^{\prime},\sqrt{2\left( \varepsilon_z + \bar \Phi \left(z^{\prime}\right) \right)}\right)-
\mathcal{T}\left(z,\sqrt{2\left( \varepsilon_z + \bar \Phi \left(z\right) \right)}
\right) \right)\right)\\
&+
\exp{\left(\left(\nu+\mathrm i \omega\right)
\left(
\mathcal{T}\left(z^{\prime},-\sqrt{2\left( \varepsilon_z + \bar \Phi \left(z^{\prime}\right) \right)}\right)-
\mathcal{T}\left(z,\sqrt{2\left( \varepsilon_z + \bar \Phi \left(z\right) \right)}\right) \right)\right)}\\
&\left.J_0\left(k \sqrt{2\left(\varepsilon-\varepsilon_z\right)}
\left(
\mathcal{T}\left(z^{\prime},-\sqrt{2\left( \varepsilon_z + \bar \Phi \left(z^{\prime}\right) \right)}\right)-
\mathcal{T}\left(z,\sqrt{2\left( \varepsilon_z + \bar \Phi \left(z\right) \right)}
\right) \right)\right)\right)
\mathrm{d}\varepsilon_z 
\end{aligned};
\end{align}
when $z^{\prime} > z$,
\begin{align}
\begin{aligned}
&K_0 (\nu+\mathrm{i}\omega,k,z,z^{\prime},\varepsilon)\\
=&\int_{-\bar\Phi\left(z\right)}^{\varepsilon} 
\frac{1}{4\sqrt{
2\left(\varepsilon+\bar \Phi \left(z\right)\right)
\left(\varepsilon_z+\bar \Phi \left(z\right)\right)
\left(\varepsilon_z+\bar \Phi \left(z^{\prime}\right)\right)}}\\  
&\left(\exp{\left(\left(\nu+\mathrm i \omega\right)
\left(
\mathcal{T}\left(z^{\prime},-\sqrt{2\left( \varepsilon_z + \bar \Phi \left(z^{\prime}\right) \right)}\right)-
\mathcal{T}\left(z,\sqrt{2\left( \varepsilon_z + \bar \Phi \left(z\right) \right)}\right) \right)\right)}\right.\\
&J_0\left(k \sqrt{2\left(\varepsilon-\varepsilon_z\right)}
\left(
\mathcal{T}\left(z^{\prime},-\sqrt{2\left( \varepsilon_z + \bar \Phi \left(z^{\prime}\right) \right)}\right)-
\mathcal{T}\left(z,\sqrt{2\left( \varepsilon_z + \bar \Phi \left(z\right) \right)}
\right) \right)\right)\\
&+
\exp{\left(\left(\nu+\mathrm i \omega\right)
\left(
\mathcal{T}\left(z^{\prime},-\sqrt{2\left( \varepsilon_z + \bar \Phi \left(z^{\prime}\right) \right)}\right)-
\mathcal{T}\left(z,-\sqrt{2\left( \varepsilon_z + \bar \Phi \left(z\right) \right)}\right) \right)\right)}\\
&\left.J_0\left(k \sqrt{2\left(\varepsilon-\varepsilon_z\right)}
\left(
\mathcal{T}\left(z^{\prime},-\sqrt{2\left( \varepsilon_z + \bar \Phi \left(z^{\prime}\right) \right)}\right)-
\mathcal{T}\left(z,-\sqrt{2\left( \varepsilon_z + \bar \Phi \left(z\right) \right)}
\right) \right)\right)\right)
\mathrm{d}\varepsilon_z 
\end{aligned}.
\end{align}

\pagebreak

\section{Appendix B. Poisson equation}
\label{section Poisson equation}
\setcounter{equation}{0}
\renewcommand{\theequation}{B.\arabic{equation}} 

Substituting the solution of kinetic equation \eqref{solution of kinetic equation} into $\displaystyle{\int \bar f \,\delta\!\underline{\tilde{f}} \,\mathrm{d}^3 v}$, the first part gives
\begin{align}
\begin{aligned}
& \int_{-\infty}^{+\infty}\int_{-\infty}^{+\infty}\int_{-\infty}^{+\infty} \int_{0}^{+\infty}
\frac{1}{\left(2 \pi\right)^{3/2}}
\exp\left(-\frac{1}{2}\left(v_x^2+v_y^2+v_z^2\right)+\bar{\Phi}\left(z\right)\right)
\delta\underline{\tilde{\Phi}}\left( Z\left(\varepsilon_z,\tau \right) \right)\\
&\delta\left(Z \left(\varepsilon_z,\tau \right)-z \right) 
\mathrm{d} Z \, \mathrm{d} v_x \, \mathrm{d} v_y \, \mathrm{d} v_z\\
=&\exp\left(\bar{\Phi}\left(z\right)\right)\delta\underline{\tilde{\Phi}}\left( z \right)
\end{aligned},
\end{align}
the second part gives
\begin{align}
\begin{aligned}
&-\left(\nu+\mathrm{i} \omega \right)
\int_{-\infty}^{+\infty} \int_{-\infty}^{+\infty} \int_{-\infty}^{+\infty} \int_{0}^{+\infty}
\int_{-\infty}^{\tau}
\frac{1}{\left(2 \pi\right)^{3/2}}\exp\left(-\frac{1}{2}\left(v_x^2+v_y^2+v_z^2\right)+\bar{\Phi}\left(z\right)\right)\\
&\exp\left(\left(\nu+\mathrm{i} \left(\omega - k_x v_x - k_y v_y\right)\right)\left(\tau^{\prime}-\tau\right) \right) \delta\underline{\tilde{\Phi}}\left(Z\left(\varepsilon_z,\tau^{\prime}\right)\right)
\delta\left(Z \left(\varepsilon_z,\tau \right)-z \right) 
\mathrm{d} \tau^{\prime} \, \mathrm{d} Z \, 
\mathrm{d} v_x \, \mathrm{d} v_y \, \mathrm{d} v_z\\
=&-\left(\nu+\mathrm{i} \omega \right) \int_{-\bar{\Phi}\left(z\right)}^{+\infty}
\frac{\exp\left(-\varepsilon_z \right)}{2\sqrt{\pi \left(\varepsilon_z+\bar \Phi \left(z\right)\right)}}
\int_{-\infty}^{+\infty} \int_{-\infty}^{\tau}
\exp\left(\left(\nu+\mathrm{i}\omega\right)\left(\tau^{\prime}-\tau\right) -\frac{k^2\left(\tau^{\prime}-\tau\right)^2}{2}\right)\\ &\delta\underline{\tilde{\Phi}}\left(Z\left(\varepsilon_z,\tau^{\prime}\right)\right)
\left(\delta \left(\tau-\mathcal{T}\left(z,\sqrt{2\left( \varepsilon_z + \bar \Phi \left(z\right) \right)}\right)\right)+\delta\left(\tau-\mathcal{T}\left(z,-\sqrt{2\left( \varepsilon_z + \bar \Phi \left(z\right) \right)}\right)\right)\right)\\
&\mathrm{d} \tau^{\prime} \, \mathrm{d} \tau \,\mathrm{d} \varepsilon_z\\
=&-(\nu+\mathrm{i}\omega) \int_{0}^{+\infty} K_1(\nu+\mathrm{i}\omega,k,z,z^{\prime}) \delta\underline{\tilde{\Phi}} (z^{\prime}) \mathrm{d} z^{\prime}
\end{aligned},
\end{align} 
and the third part gives
\begin{align}
\begin{aligned}
& \nu \int_{-\infty}^{+\infty}\int_{-\infty}^{+\infty}\int_{-\infty}^{+\infty} 
\int_{0}^{+\infty} \int_{-\infty}^{\tau} 
\frac{1}{\left(2\pi\right)^{3/2}}
\exp\left(-\frac{1}{2}\left(v_x^2+v_y^2+v_z^2\right)+\bar{\Phi}\left(z\right)\right) \\
& \exp\left(\left(\nu+\mathrm{i} \left(\omega - k_x v_x - k_y v_y\right)\right)\left(\tau^{\prime}-\tau\right) \right) \,\delta\underline{\tilde{g}} \left( Z\left(\varepsilon_z,\tau^{\prime}\right),\frac{1}{2}\left( v_x^2+v_y^2 \right)+\varepsilon_z \right) 
\delta\left(Z \left(\varepsilon_z,\tau \right)-z \right) \\
& \mathrm{d} \tau^{\prime}\, \mathrm{d} Z \, \mathrm{d} v_x \, \mathrm{d} v_y \, \mathrm{d} v_z \\
=&\nu \int_{-\bar{\Phi}\left(z\right)}^{+\infty} \int_{0}^{+\infty} 
K_2(\nu+\mathrm{i}\omega,k,z,z^{\prime},\varepsilon)
\delta\underline{\tilde{g}} (z^{\prime},\varepsilon) 
\mathrm{d} z^{\prime} \mathrm{d} \varepsilon
\end{aligned}.
\end{align}

When $z^{\prime} \leq z$,
\begin{align}
\begin{aligned}
&K_1 (\nu+\mathrm{i}\omega,k,z,z^{\prime})\\
=& \int_{-\bar{\Phi}\left(z^{\prime}\right)}^{+\infty}
\frac{\exp\left(-\varepsilon_z \right)}
{2\sqrt{2 \pi 
\left(\varepsilon_z+\bar \Phi \left(z\right)\right)
\left(\varepsilon_z+\bar \Phi \left(z^{\prime}\right)\right)}}
\\
&\left(
\exp\left(\left(\nu+\mathrm{i}\omega\right)
\left(
\mathcal{T}\left(z^{\prime},\sqrt{2\left( \varepsilon_z + \bar \Phi \left(z^{\prime}\right) \right)}\right)
-\mathcal{T}\left(z,\sqrt{2\left( \varepsilon_z + \bar \Phi \left(z\right) \right)}\right)\right)\right.\right.\\ 
&\left.-\frac{k^2}{2}
\left(
\mathcal{T}\left(z^{\prime},\sqrt{2\left(\varepsilon_z + \bar \Phi \left(z^{\prime}\right) \right)}\right)
-\mathcal{T}\left(z,\sqrt{2\left(\varepsilon_z + \bar \Phi \left(z\right) \right)}\right)\right)^2\right)\\
&+\exp\left(\left(\nu+\mathrm{i}\omega\right)
\left(
\mathcal{T}\left(z^{\prime},-\sqrt{2\left( \varepsilon_z + \bar \Phi \left(z^{\prime}\right) \right)}\right)
-\mathcal{T}\left(z,\sqrt{2\left( \varepsilon_z + \bar \Phi \left(z\right) \right)}\right)\right)\right.\\ 
&\left.\left.
-\frac{k^2}{2}
\left(
\mathcal{T}\left(z^{\prime},-\sqrt{2\left( \varepsilon_z + \bar \Phi \left(z^{\prime}\right) \right)}\right)
-\mathcal{T}\left(z,\sqrt{2\left( \varepsilon_z + \bar \Phi \left(z\right) \right)}\right)\right)^2\right)\right)
\mathrm{d} \varepsilon_z
\end{aligned};
\end{align}
when $z^{\prime} > z$,
\begin{align}
\begin{aligned}
&K_1 (\nu+\mathrm{i}\omega,k,z,z^{\prime})\\
=& \int_{-\bar{\Phi}\left(z\right)}^{+\infty}
\frac{\exp\left(-\varepsilon_z \right)}
{2\sqrt{2 \pi 
\left(\varepsilon_z+\bar \Phi \left(z\right)\right)
\left(\varepsilon_z+\bar \Phi \left(z^{\prime}\right)\right)}}
\\
&\left(\exp\left(\left(\nu+\mathrm{i}\omega\right)
\left(
\mathcal{T}\left(z^{\prime},-\sqrt{2\left( \varepsilon_z + \bar \Phi \left(z^{\prime}\right) \right)}\right)
-\mathcal{T}\left(z,\sqrt{2\left( \varepsilon_z + \bar \Phi \left(z\right) \right)}\right)\right)\right.\right.\\ 
&\left.-\frac{k^2}{2}
\left(
\mathcal{T}\left(z^{\prime},-\sqrt{2\left( \varepsilon_z + \bar \Phi \left(z^{\prime}\right) \right)}\right)
-\mathcal{T}\left(z,\sqrt{2\left( \varepsilon_z + \bar \Phi \left(z\right) \right)}\right)\right)^2\right)\\
&+\exp\left(\left(\nu+\mathrm{i}\omega\right)
\left(
\mathcal{T}\left(z^{\prime},-\sqrt{2\left( \varepsilon_z + \bar \Phi \left(z^{\prime}\right) \right)}\right)
-\mathcal{T}\left(z,-\sqrt{2\left( \varepsilon_z + \bar \Phi \left(z\right) \right)}\right)\right)\right.\\ 
&\left.\left.-\frac{k^2}{2}
\left(
\mathcal{T}\left(z^{\prime},-\sqrt{2\left( \varepsilon_z + \bar \Phi \left(z^{\prime}\right) \right)}\right)
-\mathcal{T}\left(z,-\sqrt{2\left( \varepsilon_z + \bar \Phi \left(z\right) \right)}\right)\right)^2\right)\right)
\mathrm{d} \varepsilon_z
\end{aligned}.
\end{align}

The function $K_2$ follows
\begin{align}
K_2 (\nu+\mathrm{i}\omega,k,z,z^{\prime},\varepsilon)=
2\exp\left(-\varepsilon \right)
\sqrt{\frac{\varepsilon+\bar{\Phi}\left(z\right)} {\pi}}
K_0 (\nu+\mathrm{i}\omega,k,z,z^{\prime},\varepsilon).
\end{align}

\pagebreak

\section*{Figures}
\iffigures

\begin{figure}[h!]
	\centering
	\includegraphics[width=0.8\textwidth]{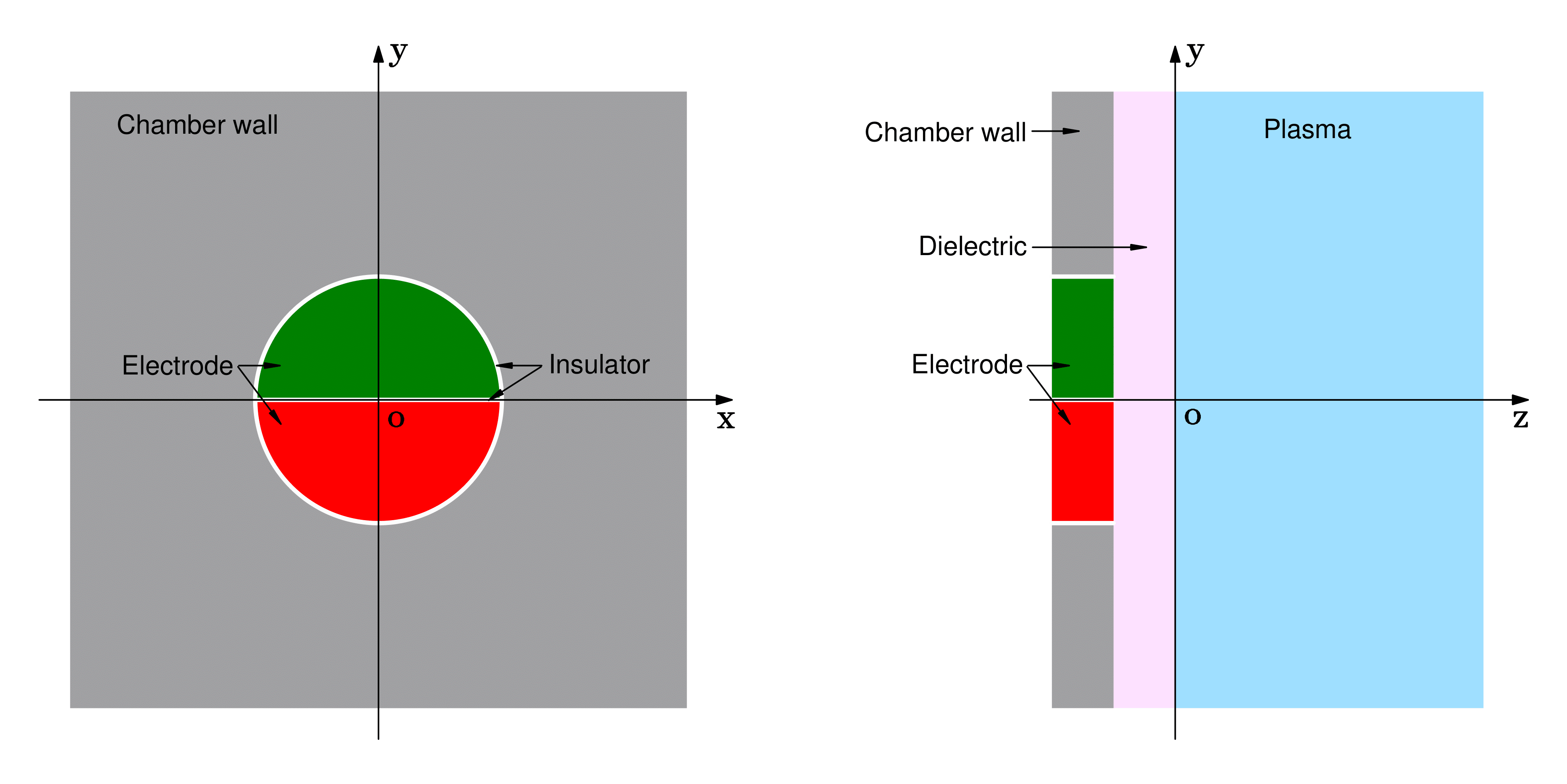} 
	\caption{Idealized planar Multipole Resonance Probe. Two semi-disc electrodes are flatly integrated into the chamber wall. A thin dielectric layer covers the electrodes and chamber wall.}
	\label{Ideal pMRP}
\end{figure}

\begin{figure}[h!]
	\centering
	\subfigure[Ion density and electron density]{
	\includegraphics[width=0.48\textwidth]{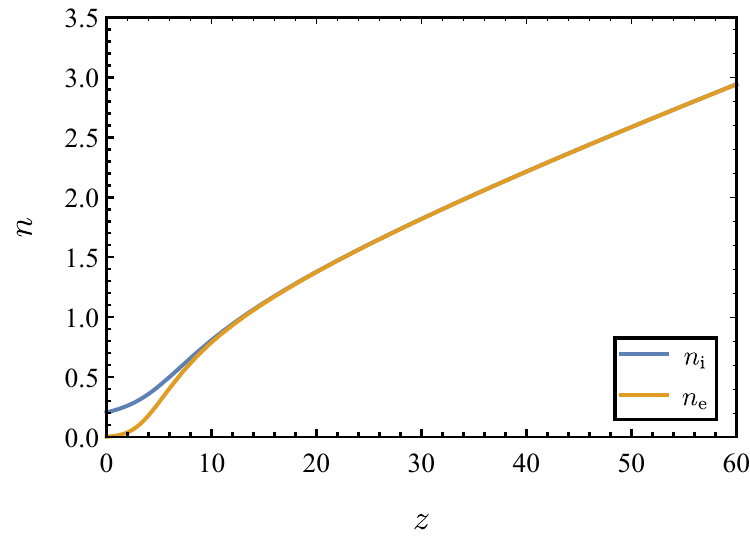}}
	\hspace{1mm}
	\subfigure[Potential]{
	\includegraphics[width=0.48\textwidth]{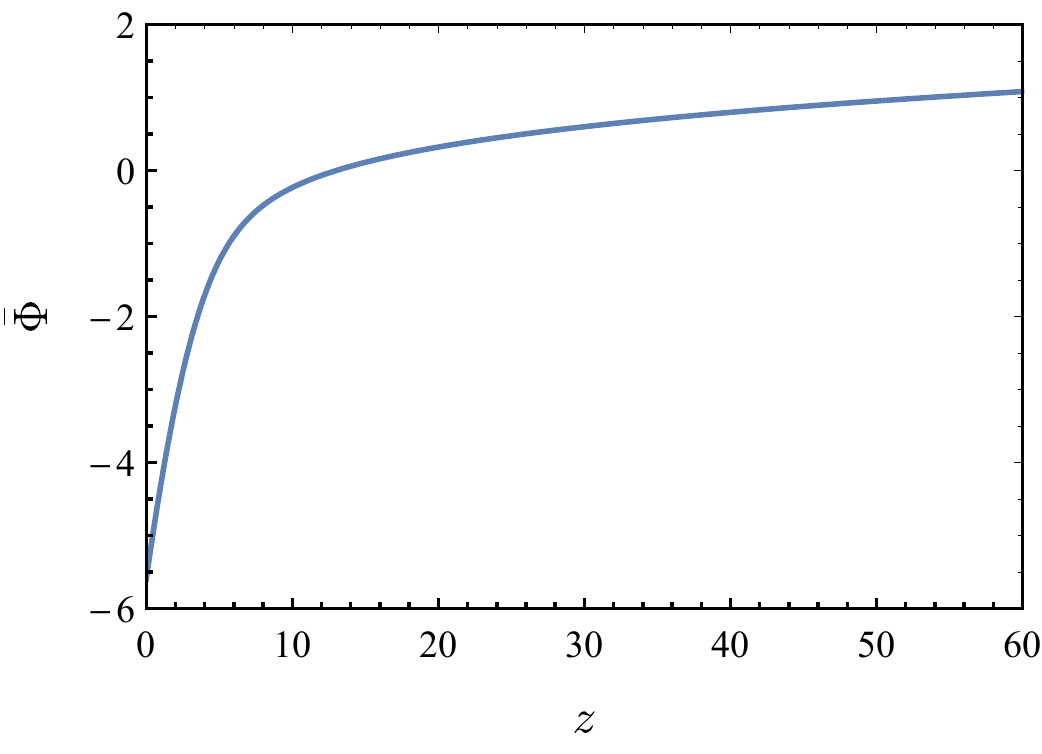}}	
	\caption{Equilibrium distribution in an argon plasma at $10\,\mathrm{Pa}$: $\hat{n}=10^{16}\,\mathrm{m}^{-3}$, $T_\mathrm{e}=2\,\mathrm{eV}$, 	 $\hat{\lambda}_\mathrm{D}=0.11\,\mathrm{mm}$.}
	\label{Sheath}
\end{figure}

\begin{figure}[h!]
	\centering
	\includegraphics[width=0.9\textwidth]{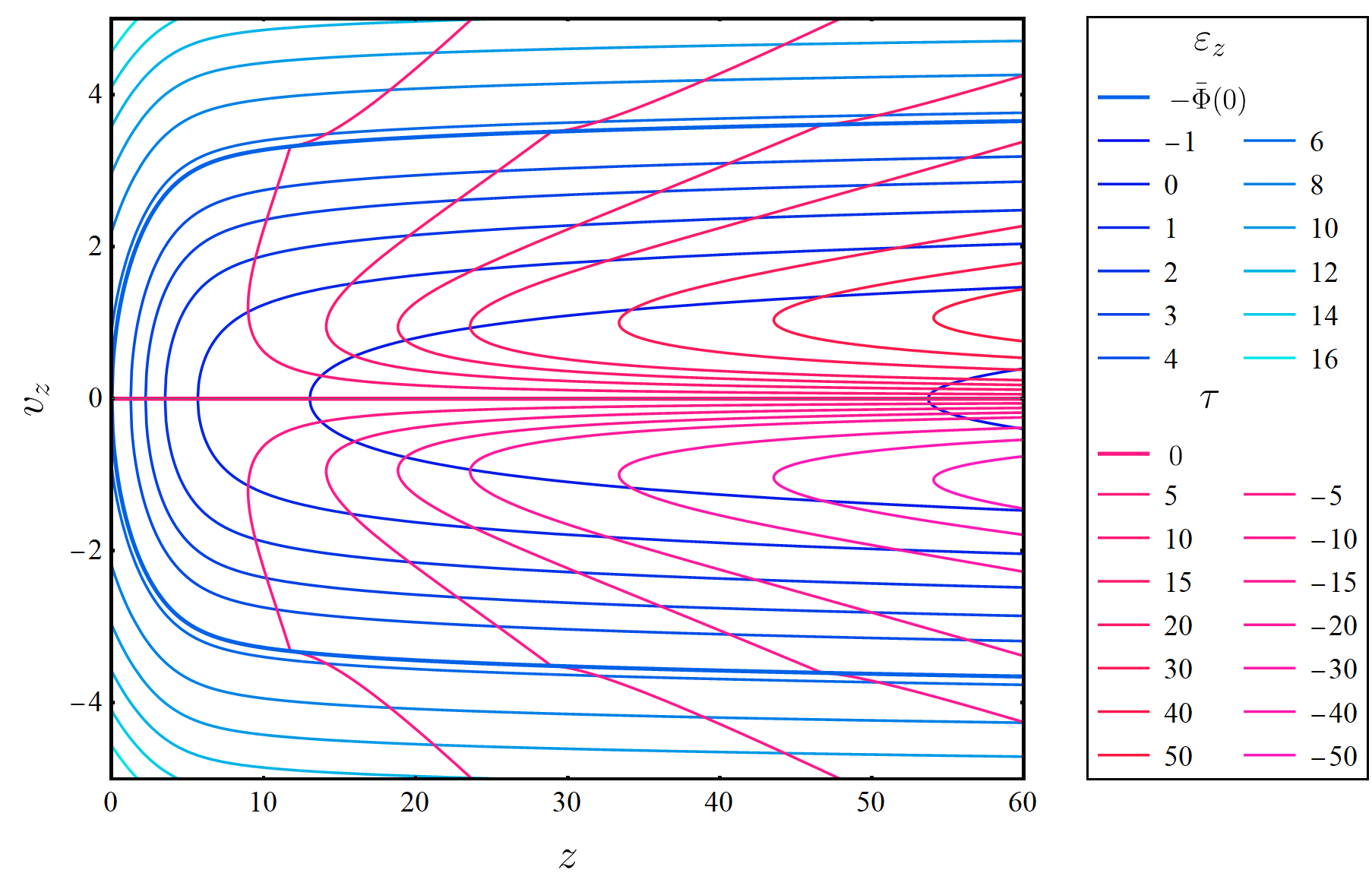} 
	\caption{Unperturbed trajectory. The blue curve shows the variation of $(z,v_z)$ at a certain $\varepsilon_z$. The red curve shows the temporal parameter $\tau$ which always takes a turning when crossing the critical trajectory $(\varepsilon_z=-\bar\Phi(0))$.}
	\label{Unperturbed trajectory}
\end{figure}

\begin{figure}[h!]
	\centering
	\includegraphics[width=0.48\textwidth]{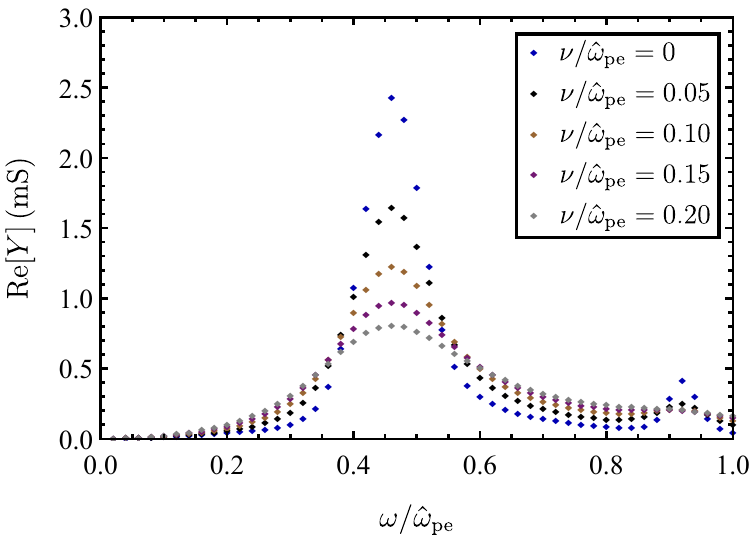} 
	\caption{Kinetic spectra of the idealized pMRP for different collision frequencies: $\hat{\omega}_\mathrm{pe}=5.64\,\mathrm{GHz}$.}
	\label{Spectra}
\end{figure}

\begin{figure}[h!]
	\centering
	\subfigure[Amplitude]{
	\includegraphics[width=0.48\textwidth]{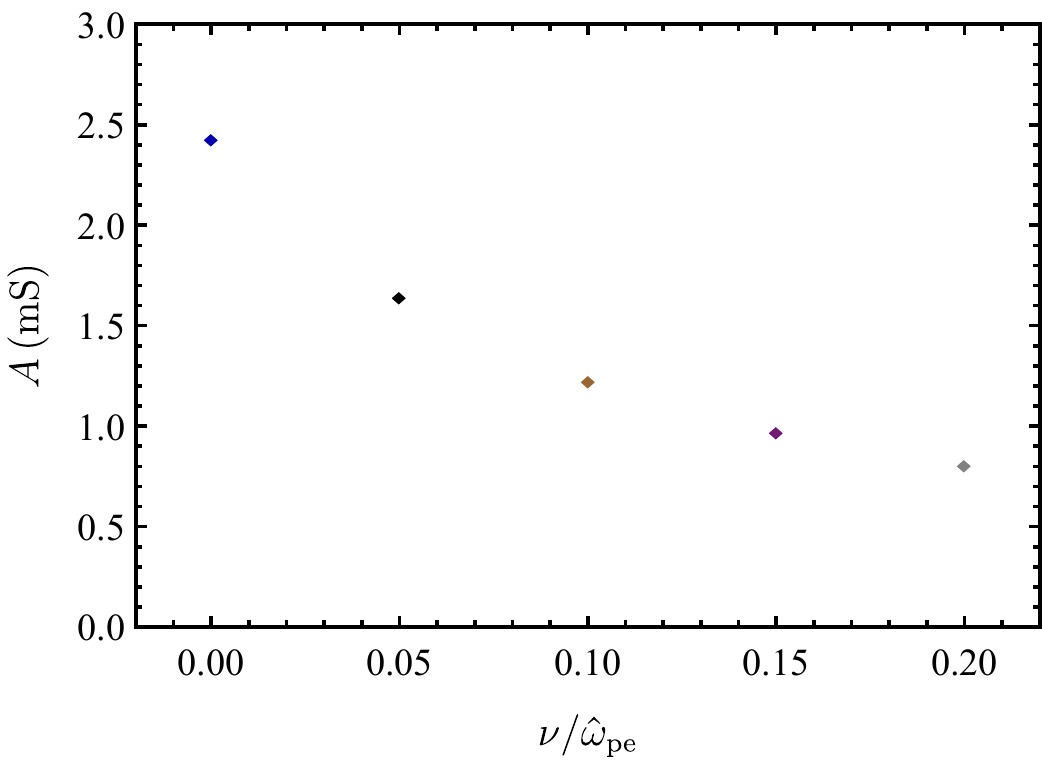}}
	\hspace{1mm}
	\subfigure[Half-width]{
	\includegraphics[width=0.48\textwidth]{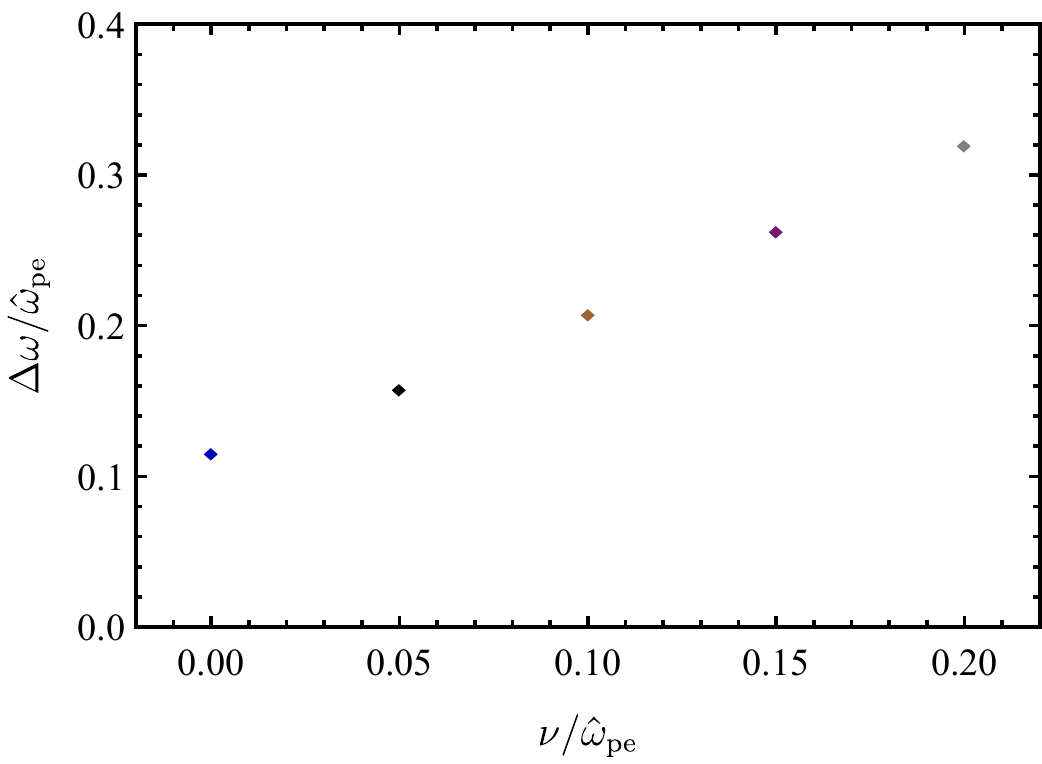}}	
	\caption{Kinetic spectra of the idealized pMRP dependent on the collision frequency: $\hat{\omega}_\mathrm{pe}=5.64\,\mathrm{GHz}$.}
	\label{Spectra12}
\end{figure}

\end{document}